\documentclass[12pt]{article}
\usepackage{amsfonts,amssymb}
 \newcommand{\End}{\nonumber\\ }
 \newcommand\Real{ {\mathbb R} }
 \newcommand{\Frac}[2]{{\textstyle{\frac{#1}{#2}}}}
 \newcommand\Half{\Frac{1}{2}}
 \newcommand\gexpect{ {{\mathbb{E}}_G} }
 \newcommand{\ZeT}{[0,T]}
 \newcommand{\Ssf}[1]{{{\mathcal F}(#1)}}
 \newcommand{\Ng}[1]{|#1|_{{}_G}}
 \newcommand{\Gp}[1]{\epsilon_{#1}}
 \newcommand{\tN}[1]{t^{{}_{[N]}}_{#1}}
 \newcommand\bint{\int_{\mathcal B}}
 \newcommand\meq{=_{\mu}}
 \newcommand\mlim{\mathop{\mu-\lim}\limits}
 \newcommand{\Proofskip}{\vskip 0.2in}
\newtheorem{definition}{Definition}[section]
\newtheorem{prop}[definition]{Proposition}
\newtheorem{theorem}[definition]{Theorem}

\newtheorem{example}[definition]{Example}
\newtheorem{cor}[definition]{Corollary}
 
\begin{document}
 \begin{center}
 {\large A Feynman-Kac Formula for Anticommuting Brownian Motion}\\
 \ \\
 \begin{tabular}{ccc}
   Steven Leppard\footnotemark[1] & and          &  Alice Rogers\footnotemark[2]\\
  \end{tabular}
   \\
  \ \\
  Department of Mathematics      \\
  King's College             \\
  Strand, London  WC2R 2LS         \\
 \end{center}
 \footnotetext[1]{current address: Enron Europe Research Group,
 Enron House, 40 Grosvenor Place, London SW1X 7EN, steven.leppard@enron.com}
 \footnotetext[2]{alice.rogers@kcl.ac.uk}
 \vskip0.2in
 \begin{abstract}
Motivated by application to quantum physics, anticommuting
analogues of Wiener measure and Brownian motion are constructed.
The corresponding It\^o integrals are defined and the existence and
uniqueness of solutions to a class of stochastic differential
equations is established.  This machinery is used to provide a
Feynman-Kac formula for a class of Hamiltonians. Several specific
examples are considered.
 \end{abstract}
\section{Introduction}
Anticommuting variables occur in physics when either a
supersymmetry or a BRST symmetry occurs. In the first place such
variables occur as the parameters of each of these two kinds of
symmetry transformations, but they also occur when the operators
of the quantized theory are represented by differential operators
on function spaces: the presence of canonical {\em
anti}-commutation relations means that the functions involved are
functions of anticommuting variables, an idea which goes back
originally to work of Martin \cite{Martin} and ideas of Schwinger
\cite{Schwin}, and was extensively developed by Berezin
\cite{Berezi1966} and by De Witt \cite{Dewitt1984}. Anticommuting
variables are not used to model physical quantities directly;
their use is motivated by the algebraic properties of the function
spaces of these variables. In application to physics, results
which are real or complex numbers emerge after what has become
known as  Berezin integration (defined by equation (\ref{berint})
in Section~\ref{GrassVarSec}) which essentially takes a trace. The
approach using anticommuting variables is particularly useful in
the context of supersymmetry and BRST symmetry because bose and
fermi (or ghost) degrees of freedom, which are related by symmetry
transformations, are both handled in the same way.

Path integral quantization in this approach has been developed in
terms of limits of time-slicing by a number of authors, starting
from the work of Martin \cite{Martin} with further work by, among
others, Marinov \cite{Marino}. A clear account of this use of
Grassmann variables in fermionic quantization is given by Swanson
\cite{Swanso}.

In this paper we investigate a more rigorous, mathematical
approach to the path integral  quantization  of ghost Hamiltonians
by developing anticommuting analogues to various constructions in
probability theory (such as Brownian motion and stochastic
calculus) and applying these objects to establish a Feynman-Kac
formula for a wide family of ghost Hamiltonians of the kind which
occur when quantizing systems in the BRST approach. These
anticommuting analogues  are constructed in close parallel to
their classical commuting counterparts, so that the two may
readily be combined to give a `super' theory in a geometric
setting. The anticommuting Brownian motion developed here is
distinct from that developed by one of the authors for fermionic
quantization \cite{GBM,PIAVS}, essentially because these two
classes of theory  have distinct free Hamiltonians.

Other approaches to quantization of fermionic and ghost degrees of
freedom have been considered by several authors: it is not
possible to give a full list, but examples are the work of  Gaveau
and Schulman \cite{GavSch}, Applebaum and Hudson \cite{AppHud} and
Hudson and Lindsay \cite{HudLin}, and Kupsch \cite{Kupsch1987}.
Closest to the work presented here is the work of Barnett,
Streater and Wilde \cite{BarStrWil1982,BarStrWil1983} and of
Hasagawa and Streater \cite{HasStr}, as will be discussed in more
detail in Section~\ref{ABMsec}.
\section{Anticommuting variables}
\label{GrassVarSec}
In this section we briefly describe the space of anticommuting
variables from which our processes are built, together with the key
features of the analysis of functions of such variables. Further
details may be found in \cite{PIAVS}. The approach taken, using
Grassmann algebras, is more concrete and more particular  than
strictly necessary; a more abstract  approach is possible, which
would be more mathematically economical and elegant, but would not
relate in so direct a way to the standard methods of stochastic
calculus.

The basic anticommuting algebra used is the real Grassmann algebra
with an infinite number of generators; this algebra, which is
denoted $\Real_{S}$, is a super algebra with
 $\Real_{S} := \Real_{S,0} \oplus \Real_{S,1}$
where $\Real_{S,0}$ is the even part, consisting of elements which
are a linear combination of terms each containing a product of even
numbers of the anticommuting generators, while $\Real_{S,1}$ is the
odd part. We will normally consider homogeneous elements, that is
elements $A$ which are either even or odd, with  parity denoted by
$\Gp{A}$ so that $\Gp{A}=i$ if $A$ is in $\Real_{S,i},i=0,1$. The
algebra $\Real_{s}$ is supercommutative, that is
 $A B=(-1)^{\Gp{A}\Gp{B}}B A$,
so that in particular
 $\alpha \beta = - \beta \alpha$
if and only if both $\alpha$ and $\beta$ are both odd. We shall not
need to be concerned with analysis on this space directly, and so
do not need to specify any norm. Our use of the space will be
purely algebraic.

The functions  with which we shall principally be concerned,
because of their r\^ole in ghost quantization, have as domain the
space $\Real^{0,m}_{S} := (\Real_{S,1})^{m}$. A typical element of
this space is
 $\eta := (\eta^{1},\ldots,\eta^{m}).$ (It will be assumed that
$m$ is an even number in this paper, although in other contexts
this is not necessarily the case.) We will consider functions on
this space which are {\em supersmooth} \cite{Dewitt1984,GTSM},
that is (in this simple context where we consider purely
anticommuting variables) multinomials in  the anticommuting
variables. These may be written in a  standard form if we
introduce multi-index notation: let $M_n$ denote the set of all
multi-indices of the form
 $\mu := \mu_{1} \ldots \mu_{k}$
with  $1 \leq \mu_{1} < \ldots < \mu_{k} \leq m$ together with the
empty multi-index {\small{$\emptyset$}}; also let $|\mu|$
denote the length of the multi-index $\mu$,
 $\eta^\emptyset := 1$ (the unit of $\Real_S$) and
 $\eta^\mu := 1 \eta^{\mu_1} \ldots \eta^{\mu_{|\mu|}}$.
A supersmooth function is then a function $F$ of the form
 \begin{eqnarray}
  F:\Real^{0,m}_{S} & \longrightarrow & \Real_{S} \nonumber \\
  (\eta^{1},\ldots,\eta^{m}) & \mapsto &
  \sum_{\mu \in M_{m}} F_{\mu} \eta^{\mu} \label{susmfn}
 \end{eqnarray}
where the coefficients $F_{\mu}$ are real or complex numbers.

Differentiation of multinomial functions of anticommuting variables
is defined by linearity together with the rule
 \begin{eqnarray}
  \frac{\partial \eta^{\mu}}{\partial \eta^{j}} &=&
  \left\{
  \begin{array}{cl}
 (-1)^{\ell-1} \eta^{\mu_{1}} \ldots   \widehat{\eta^{\ell}} \ldots \eta^{\mu_{|\mu|}},
   & \mbox{ if $j=\mu_{\ell}$ for some $\ell$, $1 \leq \ell \leq |\mu|$,} \\
 0 & \mbox{otherwise,}
  \end{array}
  \right.\End
 \label{graddiff}
 \end{eqnarray}
where  $\,{\widehat{}}\,$  indicates an omitted factor.

Functions of anticommuting variables obey the following Taylor
theorem, which can be proved as in the classical case.
 \begin{theorem}
If $F$ is a supersmooth function on $\Real^{0,m}_{S}$ and
$\xi,\eta$ are
 elements of $\Real^{0,m}_{S}$,
 \begin{eqnarray}
  F(\xi+\eta)-F(\xi) &=& \eta^{a_{1}}\partial_{a_{1}}F(\xi)
  + \frac{1}{2!}\eta^{a_{2}}\eta^{a_{1}}\partial_{a_{1}}
  \partial_{a_{2}}F(\xi) + \ldots \nonumber \\
 &&+ \frac{1}{(n-1)!}\eta^{a_{n-1}} \ldots \eta^{a_{1}}
  \partial_{a_{1}} \ldots \partial_{a_{n-1}} F(\xi) \nonumber \\
 &&+ \int_{0}^{1} \frac{(1-t)^{n-1}}{(n-1)!}
  \eta^{a_{n}} \ldots \eta^{a_{1}}
  \partial_{a_{1}} \ldots \partial_{a_{n}} F(\xi + t\eta) dt. \nonumber \\
 \label{taylor}
 \end{eqnarray}
(Here and later the summation convention for repeated indices is used.) If
the number of terms $n$ is greater than the number of anticommuting
variables $m$ this takes the simpler form
 \begin{equation}
  F(\xi+\eta) = \sum_{\mu \in M_n} \eta^{\mu} \partial_{\tilde{\mu}}F(\xi)
 \end{equation}
where
 $\partial_{\tilde{\mu}}=\partial_{\mu_{|\mu|}}\dots\partial_{\mu_1}$.
 \end{theorem}

Integration of functions of these anticommuting variables is
defined algebraically by the Berezin rule:
 \begin{equation}
  \bint d^m\eta \, F(\eta) = F_{1 \ldots m},
  \label{berint}
 \end{equation}
where $F(\eta)= \sum_{\mu \in M_{m}} F_{\mu} \eta^{\mu}$ as in
(\ref{susmfn}), so that $F_{1 \ldots m}$ is the coefficient of the
highest order term.

The space of supersmooth functions of $m$ anticommuting variables
will be denoted $\Ssf{m}$, and is a $2^m$-di\-mensional vector
space. A norm on this space is defined by
 \begin{equation}\label{NORMdef}
  \Ng{F}= \sum_{\mu\in M_n} |F_{\mu}|
 \end{equation}
where again $F(\eta)= \sum_{\mu \in M_{m}} F_{\mu} \eta^{\mu}$ as
in (\ref{susmfn}). This norm has the Banach algebra property
 \begin{equation}\label{BAeq}
  \Ng{FG} \leq \Ng{F} \Ng{G}.
 \end{equation}
Any linear operator $K$ on this space has integral kernel taking
$\Real^{0,m}_{S}\times\Real^{0,m}_{S} $ into $\Real_{S}$ defined by
\begin{equation}\label{IKeq}
  K f(\theta) = \bint d^m\theta \, K(\eta,\theta) f(\theta).
\end{equation}

\section{Anticommuting probability and stochastic processes}
\label{antiStochInt}

While the standard integral for functions of anticommuting
variables, the  Berezin integral defined in equation
(\ref{berint}), has no measure-theoretic or `limit of a sum'
aspect, it can be used to build an anticommuting analogue of
probability theory by taking the consistency conditions of the
Kolmogorov extension theory as the defining properties, as has been
carried out in \cite{GBM,PIAVS}. The key definition of
anticommuting probability space is now given. A restricted form of
the definition, sufficient for this paper, is used, with more
details and generality available in the references cited.

 \begin{definition}
 \label{GPSdef}
A {\bf $(0,m)$-anticommuting probability space of weight $w$}
consists of
 \begin{enumerate}
  \item a finite closed interval $\ZeT$ of the real line;
  \item for each finite set $B  = \{t_1, \dots,t_r \}$ with
  $0\leq t_1 < \dots <t_r \leq T$, a supersmooth function $F_{B}$ on
  $(\Real_S^{0,m})^r$ such that
   \begin{enumerate}
     \item
 \begin{equation}
 \bint d^m\theta_{1} \ldots d^m\theta_{r} \,
  F_{B}(\theta_{1},\ldots \theta_{r}) = w
 \label{probsp1}
 \end{equation}
 (where $\theta_1, \dots, \theta_r$ are each elements of $\Real_S^{0,m}$);
    \item if $B=\{t_{1},\ldots t_{r}\}$ and
    $B'=\{t_{1}, \ldots t_{r-1}\}$ then
 \begin{equation}
 \bint d^m\theta_{r} \, F_{B}(\theta_{1},\ldots \theta_{r})
  = F_{B'}(\theta_{1},\ldots \theta_{r-1}) .
   \label{probsp2}
 \end{equation}
   \end{enumerate}
  \end{enumerate}
Such a space will be denoted
 $((\Real_{S}^{0,m})^{\ZeT},\{F_{B}\},d \mu)$.
 \end{definition}
(The conditions (\ref{probsp1}) and (\ref{probsp2}) are analogous
to the consistency conditions for finite-dimension\-al
distributions.)

We can now define the notion of random variable on this space; we
cannot use conventional measure theory, but must instead build an
explicit limiting process into the definition.
 \begin{definition}
 \label{ABMmrvDefn}
A {\bf $(0,k)$-dimensional anticommuting random variable}
 \begin{equation}
  G^i := (G^i_{r},B_{r}), \ i=1,\ldots,k,
 \label{ABMmrv}
 \end{equation}
for the anticommuting probability space
 $((\Real_{S}^{0,m})^{\ZeT},\{F_{B}\},d\mu)$ consists of
 \begin{enumerate}
  \item a sequence of defining sets $B_{1}, B_{2}, \ldots$, each a finite
  subset of $\ZeT$;
  \item a sequence of supersmooth functions
   $G_{r}:(\Real_{S}^{0,m})^{|B_{r}|} \to \Real_{S}^{0,k}, \, r=1,2,\dots$ 
(with components $G^i_r, i=1, \dots,k$) such that for each
$i=1,\dots,k$ and each multinomial function $H$ of $k$ variables
the sequence
 \begin{equation}
  I_{r}(H) = \bint d^m\theta_1 \dots d^m \theta_{|B_{r}|} \,
  F_{B_{r}}(\theta_1,\ldots,\theta_{|B_r|})
  H\left(G_{r}(\theta_1,\ldots,\theta_{|B_r|})\right)
 \end{equation}
tends to a limit as $r$ tends to infinity.  (Here $|B_r|$ denotes
the number of elements in the set $B_r$.)
 \end{enumerate}
The limit of $I_{r}(H)$ is called the {\bf (anticommuting)
expectation value} of $H(G^i)$, and we write
 \begin{equation}
  \gexpect[H(G^i)] \equiv \int d\mu \, H(G^i)
  := \lim_{r \rightarrow \infty}  I_{r}(H).
 \end{equation}
The case where there  exists some finite number $M$ such that
$B_{q} = B_{M}$ for all $q > M$ is called a {\bf finitely-defined
anticommuting random variable}.
 \end{definition}
The definition of a stochastic process is analogous to the
conventional one:
 \begin{definition}
Let $A$ be an interval  contained in $\ZeT$. Then a collection
 \begin{equation}
 X:=\{ \, X_t \; | \; t \in A \, \} \label{stochpr}
 \end{equation}
of  $(0,k)$-dimensional random variables on an anticommuting
probability  space\linebreak
 $((\Real_{S}^{0,m})^{\ZeT}, \{F_{B}\},d \mu)$ is said
to be a {\bf $(0,k)$-dimensional stochastic process} on the space
$((\Real_{S}^{0,m})^{\ZeT},\{F_{B}\},d \mu)$ if for each finite
subset $A_{\alpha}$ of $A$ the collection
 $
   X:=\{ \, X_t \; | \; t \in A_{\alpha} \, \}
 $
is an anticommuting random variable on this space.
\end{definition}
In this paper we  shall be concerned with stochastic processes
which are built from solutions of stochastic differential
equations.

We end this section with some useful but rather technical
definitions starting with a notion of equality of random
variables.
 \begin{definition}
 \label{ABMmeanEquality}
If $(X^i)$ and $(Y^i)$ are two $(0,k)$-dimensional random
variables and
 \begin{equation}
  \gexpect[H(X)] = \gexpect[H(Y)],
 \end{equation}
for all multinomial functions $H$ of $k$ variables, then we say
they are {\bf $\mu$-equal}. This is written
 \begin{equation}
  X^i \meq Y^i.
 \end{equation}
 \end{definition}
The next definition defines convergence of a sequence of random
variables.
 \begin{definition}
 \label{ABMmeanCvgce}
If $X$ is a $(0,k)$-dimensional random variable,
 $X_r,r=1,2,.$ a sequence of $(0,k)$-dimensional random variables,
and
 \begin{equation}
 \lim_{r \to \infty} \Big| \gexpect[H(X_r) - H(X)] \Big|
  = 0,
 \end{equation}
for each multinomial function $H$ then we say that $X_r$ {\bf
$\mu$-converges} to $X$. This will be denoted
 \begin{equation}
  \mlim_{r \to \infty} X_r = X.
 \end{equation}
 \end{definition}
While other kinds of equality and convergence can be defined, these
forms are sufficient for the purposes of this paper since the
Feynman-Kac formula is built from expectations of anticommuting
random variables.
\section{Anticommuting Brownian motion}\label{ABMsec}
The anticommuting Brownian motion process will now be
constructed. We start by defining anticommuting Wiener space,
using finite dimensional marg\-in\-al distributions built from the
heat kernel of the `free' Hamiltonian for functions of $m$
anticommuting variables.  Recalling that we are assuming that $m$
is even, this Hamiltonian is
 \begin{equation}
 H_{F} := \Half e^{ij}\frac{\partial}{\partial \eta^i}
 \frac{\partial}{\partial \eta^j}
       \label{FHAMeq}
 \end{equation}
where the $m \times m$ matrix $e$ in block diagonal form is
 \begin{equation}
  e =
   \left( \begin{array}{ccc}
    \epsilon &            &        \\
    &        \ddots       &        \\
    &        &            \epsilon \\
   \end{array} \right)
 \label{flatham}
 \end{equation}
with
 $ \epsilon = \left( \begin{array}{cc} 0  & 1 \\
                                        -1 & 0
                    \end{array}\right).
 $
The heat kernel $e^{-H_Ft}(\eta,\eta')$ of this Hamiltonian is
 \begin{equation}
  p(\eta-\eta',t)
  := (\surd t)^m
  \exp\left(\frac{e_{ji}(\eta^i-\eta'^i)(\eta^j-\eta'^j)}{2t}\right)
 \label{heatkernel}
 \end{equation}
as may be verified by observing that $p(\eta-\eta',t)$ satisfies
the equation
 \begin{equation}
  \frac{\partial }{\partial t} p(\eta,\eta',t)
  = -H_F \, p(\eta,\eta',t)
 \end{equation}
and reduces to the Grassmann delta function
 $\delta(\eta-\eta'):= \Pi_{i=1}^{m} \left(\eta^i - \eta'{}^{i} \right)$ when $t=0$.

Anticommuting Brownian motion is now defined to be the
anticommuting stochastic process constructed from this heat kernel
in the following way:
 \begin{definition}
Anticommuting Wiener space of dimension $(0,m)$ (where $m$ is even)
 on the time interval $\ZeT$
 is the anticommuting probability space
  $((\Real_{S}^{0,m})^{\ZeT},\{F_{B}\},d \mu)$ with
 \begin{equation}
 F_{\{t_1, \dots, t_N\}}(\eta_1,\ldots,\eta_N)
 := p(\eta_1,t_1) p(\eta_2-\eta_1,t_2-t_1)
 \ldots p(\eta_N-\eta_{N-1},t_N-t_{N-1})
 \label{fdd}
 \end{equation}
for each finite set $\{t_1,\dots,t_N\}$ of real numbers for which
 $0\leq t_1 < \dots < t_N \leq T$.
 \end{definition}
It follows immediately from the semigroup property of the heat
kernel \break $p(\eta,\eta',t)$ that the finite dimensional
marginal distributions
 $F_{t_1, \dots, t_N}$ satisfy the necessary consistency condition
contained in Definition \ref{GPSdef}, and then by direct
calculation that the weight of the space is $1$.

We now define $m$-dimensional  anticommuting Brownian motion to be
the stochastic process $\beta_t$ defined by this anticommuting
probability space, so that for any supersmooth function $H$ of $mN$
anticommuting variables, where $N$ is a positive integer,
 \begin{eqnarray}
 \lefteqn{ \gexpect \left[ H(\beta^a_{t_1}, \dots, \beta^a_{t_N}) \right]} \End
  &=&
  \bint d^m \theta_1 \dots d^m\theta_N
  p(\theta_1,t_1) p(\theta_2-\theta_1,t_2-t_1)
  \End
  &&
  \qquad\dots
  p(\theta_N-\theta_{N-1},t_N-t_{N-1}) H(\theta_1, \dots,\theta_N).
 \end{eqnarray}

The following expectations, which will prove useful in subsequent
sections,  may be calculated directly from this definition.
  \begin{eqnarray}\label{SEXPeq}
   \gexpect [\beta^a_{t}] &=& 0 \End
  \gexpect  [\beta^a_{t} \beta^b_{t}]  &=& e^{ab} t \End
  \gexpect[\beta^a_{t_1} \beta^b_{t_2}] &=& e^{ab} \min(t_1,t_2) \End
  \gexpect[(\beta^a_{t_2}-\beta^a_{t_1})(\beta^b_{t_2}-\beta^b_{t_1})]
   &=& e^{ab} |t_2-t_1| .
 \end{eqnarray}
An important consequence of these results is that the process
$\beta_t$ has {\bf independent increments}:
 \begin{equation}
  \gexpect[(\beta^a_{t_2}-\beta^a_{t_{1}})
  (\beta^b_{s_2}-\beta^b_{s_1})]
  = 0
 \end{equation}
if $t_2 > t_1 \geq s_2 >s_1$.

These results show that anticommuting Brownian motion has the same
covariance as the It\^o Clifford process introduced by Barnett,
Streater and Wilde \cite{BarStrWil1982,BarStrWil1983} and further
studied by Hasagawa and Streater \cite{HasStr}. From this point of
view we are providing a concrete model of these processes, and
applying them in a novel way to path integration in ghost quantum
mechanics.

The results (\ref{SEXPeq}) can be further extended if we introduce
the notion of adapted process in close analogy with the standard
definition.
 \begin{definition}\label{ADAPdef}
A  stochastic process $F_t, t \in \ZeT$ on $m$-dimensional
anticommuting Wiener space such that for each $t$ in $\ZeT$ $F_t$
is a function of  $\{\beta_s|0 \leq s \leq t\}$ is said to be {\bf
$[0,t]$-adapted}.
 \end{definition}
(The time interval, $[0,t]$, may be omitted when the context makes
it clear.) As in the classical case, it can then be shown by direct
calculation that, if $ F_t $ is a $[0,t]$-adapted process and
 $0 \leq s < u \leq T$, then
 \begin{eqnarray}\label{ADPeq}
  \gexpect[F^a_s(\beta^b_u-\beta^b_s)] &=&    0,  \End
   \mbox{and} \quad
  \gexpect[F^a_s(\beta^b_u-\beta^b_s)(\beta^c_u-\beta^c_s)] &=&
  \gexpect[F^a_s] e^{bc}(u-s).
 \end{eqnarray}
\section{Anticommuting stochastic integrals}
As in the classical case, two kinds of integrals of anticommuting
stochastic processes will be useful, those with respect to time and
those along (anticommuting) Brownian paths. Before defining these
integrals it is useful to introduce a notation for a decreasing
sequence of partitions of the interval $[0,t], t\leq T$. For $N =
1,2,
\dots$ and fixed $t$ in $\ZeT$  the set
 $\{\tN0,\tN1, \dots, \tN{N}\}$
is a subset of $\ZeT$ with $\tN0=0$, $\tN0 < \dots < \tN{N}$,
$\tN{N}=t$ and $\Delta\tN{} \equiv \sup_{r=1..N} |\tN{r}-\tN{r-1}|
 \to 0$  as $N \to \infty$.
 \begin{definition}
 \label{ABMItoTimeInt}
The {\bf integral with respect to time} of an $n$-dimensional
adapted process $A^i_s$ is defined (when it  exists independent of
the choice of decreasing sequence of partitions) to be the process
 \begin{equation}
   \int_0^t ds \, A^i_s
  := \lim_{N \to \infty} \sum_{r=1}^N (\tN{r} - \tN{r-1}) A^i_{\tN{r-1}}.
 \end{equation}
It is clearly $[0,t]$-adapted.
 \end{definition}
The anticommuting analogue of the It\^o integral will now be
defined:
 \begin{definition}
 \label{ABMItoInt}
Suppose that $C^i_{as},i=1, \dots,n,a=1,\dots,m$ is an $n \times
m$-dimensional adapted process  on anticommuting Wiener space. Then
the {\bf It\^o integral} of the process is defined (when it exists
independent of the choice of sequence of decreasing partitions) to
be
\begin{equation}\label{IIdef}
  \int_0^t d\beta^a_s \, C^i_{as}
    := \lim_{N \to \infty} \sum_{r=1}^N (\beta^a_{t_r} -
    \beta^a_{t_{r-1}}) C^i_{a t_{r-1}}.
\end{equation}
It is clearly $[0,t]$-adapted.
\end{definition}
At this stage we do not consider necessary or sufficient conditions
on the processes $A_t,C_t$ for these integrals to exist; this
question is addressed directly for the various processes considered
in applications in later sections.

 \begin{definition}
 \label{GrassItoProcess}
An {\bf anticommuting It\^o process} or {\bf anticommuting \break
stochastic integral} is a process of the form
 \begin{equation}
  Z^i_t = Z^i_0 +
  \int_0^t ds \, A^i_s + \int_0^t d \beta^a_s \, C^i_{as},
 \end{equation}
where $A^i_s$ and $C^i_{as}$ are $[0,s]$-adapted processes.
\end{definition}

Using (\ref{SEXPeq}) and (\ref{ADPeq}) the anticommuting It\^o
isometry can be proved in close analogy to the classical case \cite{Oksend}.
 \begin{prop}\label{AIIprop} Suppose that for $i=1 \dots k$
  \[
     Z^i_t= \int_0^t d\beta^a_s C^i_{as}
  \]
 with each $Z^i$  of definite Grassmann parity. Then
 \begin{equation}
 \label{GrassItoIsomeq}
  \gexpect\left[ Z^i_t Z^j_t\right]
  = \int_0^t ds \, \gexpect[(-1)^{\Gp{Z^i}}e^{ba}
   C^i_{as} C^j_{bs}].
 \end{equation}
 \end{prop}
\section{Anticommuting stochastic differential \break equations}
In this section the anticommuting analogues of stochastic
differential equations will be considered; these are applied in the
final section to give a proof of the Feynman-Kac formula for a wide
class of Hamiltonians. No very general theory is needed, a rather
prescriptive and constructive approach is taken, motivated by the
application to path integration. We simply define a sequence of
random variables which satisfy the required stochastic differential
equation.

 \begin{theorem}
 \label{GsdeThm}
Suppose that for $i=1,\dots n$ and $a=1, \dots,m$ the functions
$A^i$ and $C^i_a$ are supersmooth functions on $\Real_S^{0,m}$
Suppose also that $\zeta_0$ is an element of $ \Real^{0,m}_{S}$.
Then there exists a unique adapted process $\zeta_t$ which
satisfies the $n$-dimensional system of anticommuting stochastic
differential equations
 \begin{equation}
 \label{gsde}
 \zeta^i_t \meq \zeta^i_0 + \int_0^t ds \, A^i_s(\zeta_s)
 + \int_0^t d\beta^a_s \, C^i_{as}(\zeta_s).
 \end{equation}
 \end{theorem}
\noindent Outline of proof:
To prove existence we construct a solution as the limit of an
inductive process. Let the sequence
 $\zeta_{t,k},k=1,2,\dots, t \in \ZeT$ of $n$-dimensional
anticommuting stochastic processes be defined by
 \begin{eqnarray}
 \zeta^i_{t,0} &=& \zeta^i_0 \End
 \zeta^i_{t,k+1} &=& \zeta^i_0 + \int_0^t ds \, A^i_s(\zeta_{s,k})
 + \int_0^t d\beta^a_s \, C^i_{as}(\zeta_{s,k}).
 \end{eqnarray}
Then,  using  the It\^o isometry Proposition \ref{AIIprop}, it may
be proved by induction that there  exists a positive constant $A$
such that (for any pair of finite subsets
 $\{t_1, \dots,t_r\}$,  $\{t'_1, \dots,t'_p\}$
 of $\ZeT$ and corresponding pair of finite sets of multi-indices
 $\mu^{[1]}, \dots, \mu^{[r]}$, $\nu^{[1]}, \dots, \nu^{[p]}$)
 \begin{eqnarray}
  && \Ng{\gexpect \Big(
  (\zeta_{t_1,k}-\zeta_{t_1,k-1} )^{\mu^{[1]}} \dots
  (\zeta_{t_r,k}-\zeta_{t_r,k-1} )^{\mu^{[r]}}
  \zeta_{t'_{1,k-1}}^{\nu^{[1]}} \dots \zeta_{t'_{p,k-1}}^{\nu^{[p]}}
  \Big)} \End
  &\leq & \frac{(A^{|\mu^{[1]}|} t)^k}{k!} \dots \frac{(A^{|\mu^{[r]}|} t)^k}{k!}
  (A^{|\nu^{[1]}| + \dots + |\nu^{[p]}|})^{k-1}.
 \end{eqnarray}
This result may be used to show that for each $t$ in $\ZeT$ and
each $\mu$ in $M_n$ the sequence
 $\Ng{\gexpect(\zeta^{\mu}_{t,k})}$ is Cauchy and hence that
$\zeta_{t,k}$ converges to an anticommuting random variable
$\zeta_t$ satisfying (\ref{gsde}).

To prove uniqueness, we suppose that  $\omega_t$ is also a solution
to (\ref{gsde}). Then, again by induction over $k$, it can be shown that
there  exists a positive constant $B$  such that
 $f_{t,k} := \sup_{\mu,\nu \in M_n, \mu \not= \emptyset}
 \Ng{\gexpect (\omega^{\mu}_t-\zeta^{\mu}_{t,k})\zeta^{\nu}_{t,k}}$
satisfies
  \begin{equation}
 0 \leq f_{t,k} \leq B \int_0^tds f_{s,k}.
 \end{equation}
and hence that, for each $t$ in $\ZeT$,
 $\lim_{k\to\infty} f_{t,k}  =  0$, so that $\omega_t \meq \zeta_t$.
 \Proofskip
The stochastic differential equation (\ref{gsde}) is often written
in differential form as
 \begin{equation}
  d\zeta^i_t =    dt \, A^i_s(\zeta_t)  +  d\beta^a_t \, C^i_{at}(\zeta_t).
 \end{equation}

In order to exploit solutions to anticommuting stochastic
differential equations to gain information about diffusions, the
following It\^o formula for stochastic integrals is essential. As
in the classical It\^o theorem, there is a second order term which
would not be present in the deterministic setting.
 \begin{theorem}
Let $X^i_t, i=1,\dots,p+q$ be a stochastic process on
anticommuting Wiener space with $X^i$ even for $i=1, \dots,p$ and
$X^i$ odd for $i=p+1, \dots p+q$, and with each $X^i$ having the
form
\begin{equation}\label{SIeq}
  X^i_t = X^i_0 + \int_0^t ds\,A^i(s,\zeta_s)
  + \int_0^t d\beta^a_s C^i_a(s,\zeta_s)
\end{equation}
where $\zeta^j, j= 1, \dots, n'$ are solutions to an
$n'$-dimensional system of anticommuting stochastic differential
equations, $\beta_t$ is $m$-dimensional anticommuting Brownian
motion, and the functions $A^i,C^i_a$ such that there exists a
positive constant $K$ for which  $\Ng{A^i(t,.)} < K, \Ng{C^i_a(t,
.)}<K$ each $t$ in $\ZeT$. Then, if $F$ is a supersmooth function
of  $p$ even and $q$ odd variables (in the sense that
 \(F(X^i) = \sum_{\mu \in M_q} F_{\mu}(X^1, \dots,X^p)
   X^{\mu_1+p}\dots X^{\mu_{|\mu|}+p} \)
with each $F_{\mu}$ a smooth function of $p$ even variables which,
together with its first and second and third derivatives, is
uniformly bounded) then
  \begin{eqnarray}\label{IFeq}
 \lefteqn{ F(X_t) } \End
 &\meq& F(X_0) + \int_0^t dX^i_s \partial_i F(X_s) \End
 && \qquad
  + \Half \int_0^t ds \, (-1)^{\Gp{X^i}} e^{ab} C^{i}_b(X_s) C^{j}_a(X_s)
  \partial_j \partial_i F(X_s). \End
  \end{eqnarray}
 \end{theorem}
\noindent Outline of proof: For each of the sequence of decreasing
partitions of $[0,t]$ we note that
 \begin{equation}\label{ISUMeq}
  F(X_t)-F(X_0) =
  \sum_{r=1}^N \Delta F_r
 \end{equation}
where $\Delta F_r= F(X_{\tN{r}}) - F(X_{\tN{r-1}})$.  Now at the
$N^{\rm th}$ approximation to the stochastic integrals $X^i_t$ we
have
\begin{eqnarray}
  \Delta F_r &=& \Delta X^i_r \partial_i F(X_{\tN{r-1}})
  + \Half \Delta X^j_r \Delta X^i_r
  \partial_i \partial_j F(X_{\tN{r-1}})\End
  && \qquad + \quad \mbox{higher order terms}
\end{eqnarray}
where
\begin{equation}
  \Delta  X^i_r = A^i(\tN{r-1}, \zeta_{\tN{r-1}}) \delta \tN{r}
   + \delta \beta^a_{\tN{r}} C^i_a(\tN{r-1}, \zeta_{\tN{r-1}}).
\end{equation}
If we now take the $k^{\rm th}$ approximation to $\zeta_t$ we can
show by induction, using the anticommuting It\^o isometry, that the
only terms in the sum (\ref{ISUMeq}) which are of order
$(\delta\tN{r})^1$ are $\Delta X^i_t \partial_i F(X_{\tN{r}})$
(coming from the first order terms in the Taylor expansion) and
 $$\Half\delta \tN{r} \, (-1)^{\Gp{X^i}} e^{ab}
 C^{i}_b(\tN{r-1}, \zeta_{\tN{r-1}})  C^{j}_a (\tN{r-1}, \zeta_{\tN{r-1}})
  \partial_j \partial_i F(X_{\tN{r}}) $$
from the second order term. All other terms are of higher order in
 $\delta\tN{r}$ and thus do not contribute to the sum in the limit
as $N$ tends to infinity.

A simple but useful special case of this theorem is the integration
by parts formula contained in the following corollary.
 \begin{cor}
The differential of the product of two stochastic integrals of the
form {\rm (\ref{SIeq})} is given by the integration by parts formula
 \begin{equation}\label{IBPeq}
  d(X^1_t X^2_t) = X^1_t dX^2_t + dX^1_t X^2_t
 + \Half (-1)^{\Gp{X^1}} e^{ab} C^1_a(t, \zeta_t) C^2_b(t, \zeta_t)\,dt.
 \end{equation}
 \end{cor}
An example of the solution of a particular stochastic differential
equation will now be described; the process which solves the
equation is the anticommuting analogue of the Ornstein-Uhlenbeck
process.
 \begin{example}{ \rm
 \label{GrassOrnUhl}
{\rm Consider the two-dimensional system of anticommuting
stoch\-astic differential equations
 \begin{equation}
       \zeta^i_t = \int_0^t ds (-r \zeta^i_s)
      + \int_0^t d\beta^a_s \, c^i_a,
 \end{equation}
where $i,a=1,2$ and $r, c^i_a$ are even constants.  This may be
solved using the same method as in the standard theory of
stochastic calculus, by applying the anticommuting form of the
It\^o integration by parts formula to the product $e^{rt}
\zeta^i_t$ obtaining
 \begin{eqnarray}
 e^{rt} \zeta^j_t
 &=&  \zeta^j_0 +  \int_0^t d( e^{rs}) \, \zeta^j_s
  +  \int_0^t d\zeta^i_t \,  e^{rs} \End
 &=& \zeta^j_0  +  r  e^{rs} \int_0^t ds \, \zeta^j_s
 +  \int_0^t ds \,(-r\zeta^i  e^{rs})
 +  \int_0^t d\beta^a_s \, c^i_a e^{rs} \End
 &=&  \zeta^j_0  +  \int_0^t d\beta^a_s \, c^i_a  e^{rs}.
 \end{eqnarray}
so that
 \begin{equation}\label{OUSeq}
  \zeta^i_t = \zeta^i_0 e^{-rt}
  + e^{-rt} \int_0^t d\beta^a_s \, c^i_a e^{rs}.
 \end{equation}
  }
 }\end{example}
\section{The anticommuting Feynman-Kac formula}
In this section we prove a Feynman-Kac formula for Hamiltonians
which are even, second-order differential operators on the space
$\Ssf{n}$ of supersmooth functions of $n$ anticommuting variables
of the form
 \begin{equation}\label{HAMeq}
  H = \Half g^{kj} \partial_j \partial_k
   + i \alpha^j \partial_j + v,
 \end{equation}
where $v$ is an even function in $\Ssf{n}$,
 $\alpha^i, i=1,\dots, n$  are odd functions and
  $g^{kj} = e^{ab} c^k_b c^j_a$ with
  $c^k_b, k=1, \dots, n, b=1,\dots, m$  even functions.
The approach taken is similar to that used for conventional,
commuting diffusions, as presented for instance in the books of
Arnold \cite{Arnold}, Friedman \cite{Friedm} and \O ksendal
\cite{Oksend}.
 \begin{theorem} \label{GFKthe}
If $H$ is a Hamiltonian of the form {\rm (\ref{HAMeq})} and $t$ is in
$\ZeT$ then for any $F$ in $\Ssf{n}$
 \begin{equation}\label{GFKeq}
  \left( e^{-Ht} F\right)(\xi)
  = \gexpect
  \left[ e^{-\int_0^t ds \, v(\zeta_s)} F(\zeta_t) \right],
 \end{equation}
where $\zeta_t$ is the anticommuting diffusion which starts from
$\xi$ and satisfies
 \begin{equation}
    d\zeta^j_t = -i \,dt \, \alpha^j(\zeta_t) + d\beta^a_t \, c^j_a(\zeta_t).
 \end{equation}
 \end{theorem}

\noindent Proof:  For $t \in \ZeT$ define the operator $U_t$ on
 $\Ssf{n}$ by
 \begin{equation}\label{OPdef}
  U_t F(\xi)
  = \gexpect \left(e^{-\int_0^t v(\zeta_s) ds} F(\zeta_t) \right).
 \end{equation}
Then, using the It\^o formula (\ref{IFeq}), we find that
 \begin{equation}
 U_t F(\xi)- F(\xi) = \int_0^t d s \, U_s H F(\xi)
 \end{equation}
so that $U_t = \exp-H t$ as required.
 \Proofskip
The first example of the application of this formula that we will
consider gives the basic path integral formula for the flat
Hamiltonian:
 \begin{example}{ \rm
Consider the Hamiltonian
 \begin{equation}
   H = \partial_1 \partial_2.
 \end{equation}
 acting on $\Ssf{2}$.
Working on two-dimensional anticommuting Wiener space, the
corresponding diffusion is the solution to
 \begin{equation}
  d\zeta^a_t = d\beta^a_t, \qquad a=1,2
 \end{equation}
starting from $\xi$. This has solution
 $\zeta_t = \xi + \beta_t$ so that
 \begin{eqnarray}
  e^{-Ht} F(\xi) &=& \gexpect \big[ F(\xi + \beta_t) \big] \End
   &=& \int d^2 \eta \,t \exp \left(\frac{\eta^1 \eta^2}{t} \right) F(\xi+ \eta) \End
   &=& \int d^2 \eta \, t \exp \left(\frac{(\eta^1-\xi^1) (\eta^2-\xi^2)}{t} \right)
    F(\eta)
\end{eqnarray}
simply reflecting the fact that anticommuting Brownian motion is
built from the heat kernel of this very Hamiltonian.
 }\end{example}
A closely related  example  gives the basic path integral formula
for the flat Hamiltonian with potential:
\begin{example}{ \rm
For the Hamiltonian
 \begin{equation}
   H =  \partial_1 \partial_2 + v
 \end{equation}
(with $v$ an even function) acting on $\Ssf{2}$
 \begin{equation}
  e^{-Ht} F(\xi) = \gexpect \left( e^{-\int_0^t ds \, v(\xi+\beta_s)}
   \big( F(\xi + \beta_t) \big) \right).
\end{equation}
 }\end{example}
The next example, which is also two dimensional, concerns the
Hamiltonian whose heat kernel gives the distribution for the
anticommuting Ornstein-Uhlenbeck process described in Example~\ref{GrassOrnUhl}.
 \begin{example}{ \rm
In the case of the Hamiltonian
 \begin{equation}\label{GrassOrnUhlHam}
 H= c^2  \partial_1 \partial_2
 + r (\eta^1 \partial_1 + \eta^2 \partial_2)
 \end{equation}
we must consider the diffusion $\zeta_t$ starting from $\xi$ and
satisfying
 \begin{equation}
  d\zeta^a_t = -r \zeta^a_t\,dt + c d\beta^a_t,
 \end{equation}
so that using (\ref{OUSeq})
 \begin{equation}
  \zeta^a_t = \xi^a e^{-rt}
  + e^{-rt} \int_0^t d\beta^a_s \, e^{rs}.
 \end{equation}
Applying the Feynman-Kac formula to four functions which form a
basis of $\Ssf{2}$, that is,
 $F_0(\eta) = 1$,
 $F_1(\eta) = \eta^1$, $F_2(\eta)= \eta^2$ and
 $F_{12}(\eta) = \eta^1\eta^2$,
we obtain
  \begin{eqnarray}
  \exp^{-Ht}F_0(\xi)  &=& \gexpect [1] =1   \End
  \exp^{-Ht}F_1(\xi)  &=&
  \gexpect \left[ \xi^1 e^{-rt}
  + e^{-rt} \int_0^t c\, d\beta^1_s \, e^{rs} \right]    \End
                      &=& e^{-rt} \xi^1       \End
  \exp^{-Ht}F_2(\xi)  &=& e^{-rt} \xi^2 \quad \mbox{and} \End
    \exp^{-Ht}F_{1}(\xi)  &=&
   \gexpect \Big[
 \left(\xi^1 e^{-rt} + e^{-rt} \int_0^t c\, d\beta^1_s \, e^{rs}\right) \End
 && \qquad \times \left(\xi^2 e^{-rt} + e^{-rt} \int_0^t c\, d\beta^2_s \, e^{rs}\right)
 \Big]    \End
 &=& e^{-2rt} \xi^1\xi^2 + \int_0^t d s \, c^2 e^{2rs}e^{-2rt} \End
 &=& e^{-2rt} \xi^1\xi^2 + \frac{c^2}{2r}(1-e^{-2rt})
 \end{eqnarray}
so that the heat kernel for this Hamiltonian is
 \begin{equation}
 e^{-Ht} (\xi,\eta) = \eta^1\eta^2
   - e^{-rt}(\xi^1 \eta^2 + \eta^1\xi^2)
   +\frac{c^2}{2r}(1-e^{-2rt}) +e^{-2rt}\xi^1\xi^2.
 \end{equation}
 }\end{example}
The next example we consider is the anticommuting Harmonic
Oscillator. This is the fundamental example in BRST quantization in
the sense that quantizing a quantum mechanical system with $k$
momenta constrained to be zero leads to a ghost Hamiltonian with
the form of the $2k$-dimensional anticommuting harmonic oscillator
\cite{HenTei,GFBFVQ}. For simplicity we consider only the two
dimensional case.
 \begin{example}\label{GrassHarmOsc}{ \rm
Consider the Hamiltonian
 \begin{equation}\label{GrassHarmOscHam}
  H = \partial_1 \partial_2 -  \eta^1 \eta^2
 \end{equation}
which leads to the anticommuting diffusion
 \begin{equation}
   \zeta_t^a = \xi^a + \beta^a_t, a=1,2.
 \end{equation}
The anticommuting Feynman-Kac formula for this diffusion is
 \begin{equation}
 \left( e^{-Ht} F \right)(\xi)
 = \gexpect\left[
  e^{\int_0^t \, ds (\xi^1+\beta_s^1) (\xi^2+\beta_s^2)}
  F(\xi+\beta_t) \right].
 \label{harmOscFK}
 \end{equation}
To evaluate this integral for finite $t$ we will use essentially
the same technique as that employed by Simon in \cite{Simon}.  To
achieve this we need to extract the kernel of the time evolution
operator from this expression, and define the analogue of
conditional expectation. Taking the definition of the expectation
with respect to anticommuting Brownian motion (\ref{harmOscFK})
becomes (we put $\Delta t := t_r - t_{r-1}$)
 \begin{eqnarray}
  \lefteqn{ \left( e^{-Ht} F \right)(\eta) } \End
 &=&
 \lim_{N \rightarrow \infty}
 \bint \, d^2\eta_1 \ldots d^2\eta_N \,
 p(\eta_1,\Delta t) p(\eta_2-\eta_1,\Delta t) \ldots
 p(\eta_N - \eta_{N-1},\Delta t)  \End
 && \times \exp \left(  \sum_{r=0}^{N-1}
 \Delta t (\eta^1+\eta_r^1) (\eta^2+\eta_r^2)\right)
  \, F(\eta+\eta_N).
 \end{eqnarray}
Making a change of variables
 $  \eta_r \mapsto \eta'_r := \eta + \eta_r $
dropping the primes and replacing $\eta_N$ by $\eta'$, we obtain
 \begin{equation}
 \left( e^{-Ht} F \right)(\eta)
 = \bint d\eta' \, \left( e^{-Ht} F \right)(\eta,\eta')F(\eta'),
 \end{equation}
where
 \begin{eqnarray}\label{PBMeq}
  \lefteqn{ \left( e^{-Ht} F \right)(\eta,\eta') = } \End
 &=& \lim_{N \rightarrow \infty} \bint d\eta_1 \ldots d\eta_{N-1} \,
  \ p(\eta_1-\eta,\Delta t) p(\eta_2-\eta_1,\Delta t) \ldots
  p(\eta_{N-1}-\eta_{N-2},\Delta t)  \End
 &&\times \  p(\eta'-\eta_{N-1},\Delta t)
 \exp \left( \sum_{r=0}^{N-1} \Delta t
 \, \eta_r^1  \eta_r^2 \right) \End
 &=& \lim_{N \rightarrow \infty} \bint d\eta_1 \ldots d\eta_{N-1} \,
  \ p(\eta_1-\eta,\Delta t) p(\eta_2-\eta_1,\Delta t) \ldots
  p(\eta_{N-1}-\eta_{N-2},\Delta t)\End
  && \times
  \ p(\eta'-\eta_{N-1},\Delta t)\frac{p(\eta-\eta',t)} {p(\eta-\eta',t)}
  \exp \left(\sum_{r=0}^{N-1} \Delta t \,
  \eta_r^1  \eta_r^2 \right) \End
 &:=& p(\eta-\eta',t) \gexpect \left[
 \exp \left( \int_0^t ds \,
 \omega_s^1  \omega_s^2 \right)
 \mid \omega_0 = \eta, \ \omega_t = \eta' \right],
 \end{eqnarray}
defining both the process $\omega_t$ (which will be called {\em
pinned anticommuting Brownian motion}), and the conditional
expectation operator.

Following Simon we use the Brownian  bridge to represent such
pinned Brownian motion processes.  The (2-dimensional) {\bf
anticommuting Brownian bridge} process starting and ending at 0,
over the time interval $[0,1]$ is defined by
 \begin{equation}
   \alpha^i_s  \ := \beta^i_s - s \beta^i_1.
  \end{equation}
In close analogy with the classical case, it may be confirmed
using (\ref{SEXPeq})  that this process has covariance
 \begin{equation}
 \gexpect\left[ \alpha_s^i \alpha_u^j \right] = e^{ij} s(1-u),
 \label{BBcovar}
 \end{equation}
for $0 \leq s \leq u \leq 1$.   This allows us to express
$\omega_t$ as
 \begin{equation}
 \omega^i(s) = \eta^i\left(1 - \frac{s}{t} \right) +
 \eta'^i \frac{s}{t} + t^{1/2} \alpha^i \left( \frac{s}{t} \right).
 \end{equation}
  Since $\int_0^t ds \, f(s/t) = t \, \int_0^1 ds' \, f(s')$, we
can restrict our attention to $\omega_s$ for $0 \leq s \leq 1$.

We now take the Fourier expansion of $\alpha(s)$,
  \begin{equation}
 \alpha^i_s =
 \sum_{r=1}^\infty \ell_r \xi^i_r f_r(s), \quad i=1,2
  \end{equation}
where $\ell_r := (r\pi)^{-1}$, $f_r(s) := \sqrt{2} \sin(r\pi s)$
and the $\xi_r$ are the anticommuting analogue of independent
Gaussian random variables, that is to say, their formal measure is
  \begin{equation}
 \prod_{r=1}^\infty
 \left( d^2\xi_r \, \exp  \xi_r^1  \xi_r^2 \right).
  \end{equation}
It can be confirmed  (as in the book of Simon \cite{Simon} for the
classical case) that this Fourier expansion for the Brownian
bridge gives the same covariance as (\ref{BBcovar}) above when
expectations are taken using this formal measure.

Pinned Brownian motion $\omega(s)$ thus has the Fourier expansion
 \begin{equation}
 \omega^i(s) = \sum_{r=1}^\infty f_r(s)
 (\gamma_r^i
  + \sqrt{t}\ell_r \xi_r^i),
 \label{pinnedBMFour}
 \end{equation}
 where $\gamma^i_r =\sqrt{2} \ell_r(\eta^i + (-1)^{r+1}\eta'^i)$.
Substituting this into the expression (\ref{PBMeq}) for the kernel
of the time evolution operator we obtain
 \begin{eqnarray}
  \lefteqn{\left( e^{-Ht} F \right)(\eta,\eta')  } \End
 &=&  p(\eta-\eta',t) \,
 \bint \left( \prod_{r=1}^\infty
 d^2\xi_r \exp \xi_r^1  \xi_r^2 \right) \End
 && \times \  \exp  \left[ \int_0^1 d s \, t \,\left(\sum_{r=1}^\infty f_r(s)
 (\gamma_r^1 + \sqrt{t}\ell_r \xi_r^1)\right)
  \left(\sum_{r=1}^\infty f_r(s)
 (\gamma_r^2 + \sqrt{t}\ell_r \xi_r^2)\right) \right]\End
 &=& p(\eta-\eta',t) \,
 \bint \left( \prod_{r=1}^\infty d\xi_r \exp \xi_r^1  \xi_r^2 \right)
  \exp  \sum_{r=1}^\infty t (\gamma_r^1 + \sqrt{t}\ell_r \xi_r^1)
  (\gamma_r^2 + \sqrt{t}\ell_r \xi_r^2) \End
 &=& p(\eta-\eta',t) \,
 \bint \left( \prod_{r=1}^\infty d\xi_r \exp \xi_r^1 \xi_r^2 \right) \End
 && \times \
  \exp \sum_{r=1}^\infty t^2 \ell_r^2
 (\gamma_r^1 t^{-1/2} \ell_r^{-1} + \xi_r^1)
 (\gamma_r^2 t^{-1/2} \ell_r^{-1} + \xi_r^2).
 \end{eqnarray}
Evaluating the Gaussian integrals we obtain
 \begin{eqnarray}
  \lefteqn{  \left( e^{-Ht}F \right)(\eta,\eta')} \End
 &=& t \prod_{r=1}^\infty (1+t^2 \ell_r^2)
  \exp \left[(\eta^1 \eta^2 + \eta'{}^1 \eta'{}2)
 \left( \frac{1}{t} + \sum_{r=1}^\infty \frac{2t \ell_r^2}{1+t^2 \ell_r^2}\right)\right]
 \End
 && \times \ \exp \left[\left(\eta^1  \eta'{}^2 + \eta'{}^1 \eta^2 \right)
 \left( \frac{1}{t} +
  \sum_{r=1}^\infty \frac{2 (-1)^r t \ell_r^2}{1+t^2 \ell_r^2}\right) \right].
 \end{eqnarray}
 Using the Weierstrass-Hadamard factorisation of $\sinh x$:
 \begin{equation}
 \sinh x = x \, \prod_{r=1}^\infty (1+\ell_r^2 x^2),
 \end{equation}
and the Mittag-Leffler expansions of $(\sinh x)^{-1}$ and $\coth
x$:
 \begin{equation} (\sinh x)^{-1} = \frac{1}{x} + \sum_{r=1}^\infty \frac{2
(-1)^r x \ell_r^2}{1+x^2 \ell_r^2}, \qquad
\coth x = \frac{1}{x} + \sum_{r=1}^\infty \frac{2 x \ell_r^2}{1+x^2 \ell_r^2}
 \end{equation}
  we finally find the kernel for the time evolution
operator to be
 \begin{eqnarray}
\lefteqn{ \left( e^{-Ht}F \right)(\eta,\eta')}\End
 &=& \sinh t\, \exp  \left[\frac{1}{ \sinh t}
 \left[ (\eta^1 \eta^2 + \eta'{}^1  \eta'{}^2)\cosh t
 - (\eta^1   \eta'{}^2 + \eta'{}^1 \eta^2) \right]\right].
 \end{eqnarray}
 } \end{example}
Finally we consider an example with quartic fermionic terms.
\begin{example}{ \rm  Consider the Hamilton\-ian
 \begin{equation} H
 = (c^2+ 2b \eta^1 \eta^2 )
 \frac{\partial^2}{\partial\eta^2 \partial\eta^1}.
 \end{equation}
Following Theorem \ref{GFKthe} we consider the stochastic
differential equation
 \begin{equation}
  \zeta_t^a = \xi^a + \int_0^t d\beta_s^a(a+\frac{b}{a}\zeta_s^1\zeta_s^2).
 \end{equation}
Without actually solving this equation it can be seen by direct
calculation (together with the anticommuting It\^o isometry
Proposition \ref{AIIprop}) that
 \begin{eqnarray}
  \gexpect [ 1 ] &=& 1, \qquad
  \gexpect [\zeta^a_t] =  \xi^a , \quad a=1,2 \End
  \mbox{and} \quad
  \gexpect[\zeta_t^1\zeta_t^2]
   &=& \xi^1\xi^2 e^{-2b t} + \frac{c^2}{2b} \left(e^{-2b t}-1\right)
 \end{eqnarray}
giving the action of $e^{-HT}$ on the four elementary functions
$1$, $\eta^1$, $\eta^2$, $\eta^1\eta^2$ to be
  \begin{eqnarray}
 \exp-H t[ 1 ] &=& 1, \qquad
  \gexpect [\eta^a_t] =  \eta^a , \quad a=1,2 \End
  \mbox{and} \quad
  \gexpect[\eta_t^1\eta_t^2]
   &=& \eta^1\eta^2 e^{-2b t}
   + \frac{c^2}{2b} \left(e^{-2b t}-1 \right)
 \end{eqnarray}
leading to the expression of the heat kernel as
 \begin{equation}
  e^{-Ht} (\eta,\xi) = \delta(\eta-\xi)
  +\frac{c^2}{2b} \left(e^{-2b t}-1 \right).
 \end{equation}

 }\end{example}
\end{document}